\documentclass[aps,nofootinbib,floatfix,showpacs]{revtex4} 
\usepackage{graphicx}
\input epsf
\newcommand{\sfig}[2]{
\includegraphics[width=#2]{#1}
        }

\def\lsim{\mathrel{\raise.3ex\hbox{$<$\kern-.75em\lower1ex\hbox{$\sim$}}}}
\def\gsim{\mathrel{\raise.3ex\hbox{$>$\kern-.75em\lower1ex\hbox{$\sim$}}}}

\def\cmm2{{\,\rm cm^{-2}}}
\def\cm2{{\,{\rm cm}^2}}
\def\cmm3{{\,{\rm cm}^{-3}}}
\def\gcmm3{{\,{\rm g\,cm^{-3}}}}

\def\fun#1#2{\lower3.6pt\vbox{\baselineskip0pt\lineskip.9pt
  \ialign{$\mathsurround=0pt#1\hfil##\hfil$\crcr#2\crcr\sim\crcr}}}

\def\be{\begin{equation}}
\def\ee{\end{equation}}
\def\bea{\begin{eqnarray}}
\def\eea{\end{eqnarray}}

\def\sigv{\langle\sigma v\rangle}

\begin{document}

\title{How Dark Matter Reionized The Universe}

\author{Alexander V.~Belikov$^{1}$ and Dan Hooper$^{1,2}$}
\affiliation{$^1$Department of Astronomy \& Astrophysics, The
University of Chicago, Chicago, IL~~60637-1433}
\affiliation{$^2$Center for Particle Astrophysics, Fermi National
Accelerator Laboratory, Batavia, IL~~60510-0500}

\date{\today}
\begin{abstract}
Although empirical evidence indicates that that the universe's gas had become ionized by redshift $z \approx 6$, the mechanism by which this transition occurred remains unclear. In this article, we explore the possibility that dark matter annihilations may have played the dominant role in this process. Energetic electrons produced in these annihilations can scatter with the cosmic microwave background to generate relatively low energy gamma rays, which ionize and heat gas far more efficiently than higher energy prompt photons. In contrast to previous studies, we find that viable dark matter candidates with electroweak scale masses can naturally provide the dominant contribution to the reionization of the universe. Intriguingly, we find that dark matter candidates capable of producing the recent cosmic ray excesses observed by PAMELA and/or ATIC are also predicted to lead to the full reionization of the universe by $z \sim 6$.
\end{abstract}
\pacs{95.35.+d; 94.20.dv; 95.85.Pw; FERMILAB-PUB-09-101-A}
\maketitle

\section{Introduction}

The universe's baryonic gas has undergone two major phase changes in cosmic history. First, nuclei and electrons combined at redshift $z\approx1100$, transforming the universe from an optically thick plasma into a gas of electrically neutral atoms. More recently, these atoms have returned to an ionized state.  Although the mechanism by which this second transition took place is not yet well understood, its is commonly suggested that the first astrophysical objects to produce significant fluxes of ultraviolet light (the minimum frequency required to ionize hydrogen), quasars and/or early stars, may have reionized the universe between approximately $6<z<20$ (for reviews, see Refs.~\cite{review1,review2,review3}). 

It is not clear, however, whether either quasars or the first stars were capable of producing enough radiation to fully reionize the universe. Previous studies~\cite{Madau:1998cd,shapiro,Fan:2001ff} have found that, unless the luminosity function of quasars favored more dim (and thus unobserved) quasars at high redshifts than in the present epoch, too few quasars would have been present at high redshifts to reionize the universe alone. While it is plausible that radiation from early stars may have lead to this transition~\cite{Gnedin:1996qr,Lu:1998ff}, the limited empirical information available regarding these objects make it difficult to draw concrete conclusions regarding their role in reionization.

Another possible source of ionizing radiation at high redshifts is the annihilation~\cite{Zhang:2006fr,Mapelli:2006ej,Ripamonti:2006gr,Ripamonti:2006gq,Mapelli:2007kr,Chuzhoy:2007fg,Natarajan:2008pk,Natarajan:2009bm} or decay~\cite{Mapelli:2006ej,Ripamonti:2006gr,Ripamonti:2006gq,Mapelli:2007kr,Zhang:2007zzh,Chen:2003gz} of dark matter particles. Here, we revisit this possibility, focusing on the case of annihilating dark matter particles with masses at or near the electroweak scale ($\sim$50-1000 GeV). We include in our calculation the evolution of the halo mass function and the effects of gas heating. We also include the impact of gamma rays produced through the inverse Compton scattering of energetic electrons with the cosmic microwave background. These inverse Compton photons are especially important for reionization, as they have considerably larger cross sections with electrons (compared to photons of higher energy), which enable them to transfer approximately $\sim10^2$ times more energy into the ionization and heating of gas than prompt gamma rays from dark matter annihilations.

In contrast to previous studies, we find that electroweak scale dark matter particles can naturally play the dominant role in the reionization of the universe. Although a dark matter particle with a $\sim$100 GeV mass and a typical thermal cross section ($\sigv \sim 3 \times 10^{-26}$ cm$^3$/s), will only lead to only approximately 1-10\% of the observed ionization, a non-thermally produced dark matter candidate, such as a 100-200 GeV wino for example, could easily provide a rate of ionizations consistent with observations (without large contributions from quasars or early stars). 

Recent observations from the cosmic ray experiments PAMELA~\cite{PAMELA} and ATIC~\cite{ATIC} have been interpreted as possible indications of dark matter annihilations taking place in the Galactic Halo~\cite{pamdm1,pamdm2}. To accomplish this, however, the halo dark matter must annihilate largely to charged leptons, and with a higher annihilation rate than would be naively predicted for a thermal relic. If one interprets the PAMELA and/or ATIC signals as the products of dark matter annihilation, this leads us to consider dark matter candidates which are naturally expected to fully reionize the universe by $z \sim 6$ without significant contributions from astrophysical sources.

\section{Reionization Of Gas With Dark Matter Annihilation Products}

In this section, we calculate the fraction of baryons that are ionized by the products of dark matter annihilations. We begin by considering he annihilation rate per volume of dark matter particles at a redshift, $z^{\prime}$, which is given by
\begin{equation}
R(z^{\prime}) = \int^{\infty}_{M_{\rm min}} \frac{dn}{dM}(M,z^{\prime})(1+z^{\prime})^3 dM \frac{\sigv}{2 m^2_X} \int \rho^2(r,M) \,4 \pi r^2 dr, 
\end{equation}
where $dn/dM$ is the differential comoving number density of dark matter halos of mass $M$, $\sigv$ is the dark matter annihilation cross section, $m_X$ is the mass of the dark matter particle, and $\rho$ is the density of dark matter in a halo as a function of the distance from the center of the halo, $r$. The second integral can be written as
\begin{equation}
\int \rho^2(r,M) 4 \pi r^2 dr = \frac{M \, \bar{\rho}(z_F)}{3} \bigg(\frac{\Omega_{X}}{\Omega_M}\bigg)^2 F(c_{200}),
\end{equation}
where $\bar{\rho}(z_F)$ is the average density of dark matter density in a halo within a radius, $r_{200}$, at which the density is 200 times larger than the cosmological average (at the time of formation):
\begin{equation}
\bar{\rho}(z_F) =  200 \, \rho_c \, \Omega_M (1+z_F)^3,
\end{equation}
where $\rho_c$ is the critical density and $z_F$ is the redshift at which the halo formed. The quantity $F(c_{200})$ is determined by the shape of the dark matter halo profile. For a halo profile of the form
\begin{equation}
\rho(r) = \frac{\rho_s}{(r/r_s)^{\alpha} [1+r/r_s]^{\beta}},
\end{equation}
we have
\begin{equation}
F(c_{200}) = \frac{c^3_{200} \int^{c_{200}}_{0} dx \, x^{2-2\alpha} (1+x)^{-2\beta}}{[\int_0^{c_{200}} dx \, x^{2-\alpha} (1+x)^{-\beta}]^2},
\end{equation}
where $c_{200} \equiv r_{200}/r_s$ is the halo concentration. Throughout this study, we will adopt a halo profile of the Navarro-Frenk-White (NFW) form ($\alpha=1$, $\beta=2$)~\cite{nfw}, with concentrations given by the analytic model of Bullock {\it et al.}~\cite{bullock}.



The number density of dark matter halos of a given mass as a function of redshift is given by 
\begin{equation}
\frac{dn}{dM}(M,z) = \frac{\rho_M}{M} \frac{\ln\sigma^{-1}(M,z)}{dM} f(\sigma^{-1}(M,z)),
\end{equation}
where $\rho_M$ is the average matter density, $\sigma(M,z)$ is the variance of the linear density field, and $f(\sigma^{-1})$ is the multiplicity function. The redshift and cosmology dependence is contained in $\sigma(M,z)$, which can be defined in terms of the matter power spectrum, $P(k)$, and top-hap function, $W(k, M) = (3/k^3 R^3)[\sin(kR) - kR \cos(kR)]$, where $R = (3M/4\pi\rho_m)^{1/3}$,
\begin{equation}
\sigma^2 (M,z) = D^2(z) \int^\infty_0  P(k) W^2(k, M) k^2dk.
\end{equation}
For determining the cold dark matter power spectrum~\cite{Bardeen}, we adopt cosmological parameters as measured by WMAP~\cite{wmap} ($\Omega_bh^2 = 0.02267$, $\Omega_c h^2 = 0.1131$, $\Omega_\Lambda = 0.726$, and $h=0.705$). 
The growth function, $D(z)$, is the linear theory growth factor, normalized to unity at $z=0$~\cite{ECF}.
We use the ellipsoidal (Sheth-Tormen) form of the multiplicity function~\cite{ST}, 
\begin{equation}
f(\sigma) = A\frac{\delta_{sc}}{\sigma}(1 + \left(\frac{\sigma^2}{a\delta^2_{sc}}\right)^p)\exp\left(\frac{a\delta^2_{sc}}{\sigma^2}\right),
\end{equation}
where $p = 0.3$, $\delta_{sc} = 1.686$ and $a = 0.75$~\cite{ST2}. We fix $A = 0.3222$ by the requirement that all of the mass lies in halos, $\int dM M \frac{dn}{dM} = \rho_mS$.
The halo mass function is most sensitive to variations in the cosmological parameters $\sigma_8$ and $n_s$. Throughout this study, we adopt two sets of values for these parameters: $n_s = 0.96, \sigma_8 = 0.812$ and $n_s = 0.986, \sigma_8 = 0.864$. The first of these sets consists of the WMAP (5 year) average values, whereas the second contains the values corresponding to the $2\sigma$ upper limit.

\begin{figure}[thbp]
    \sfig{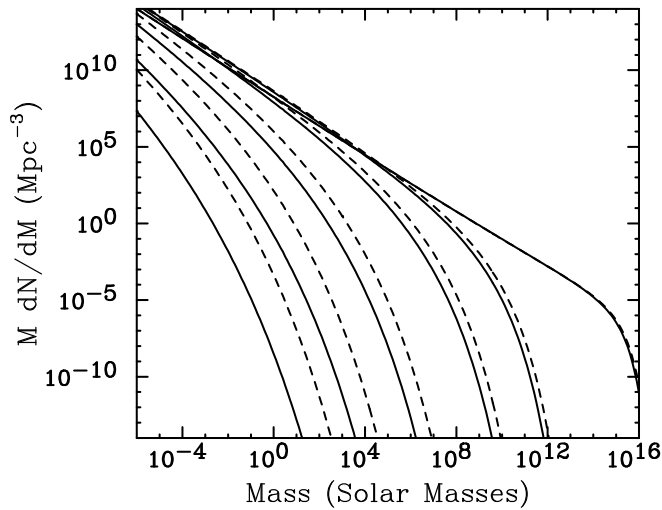}{0.49\columnwidth}
    \caption{The comoving number density of dark matter halos as a function of mass, at redshifts of 80, 60, 40, 20, 10 and 0 (from bottom-to-top). The solid (dashed) lines were calculated using $\sigma_8=0.812$ and $n_s=0.96$ ($\sigma_8=0.864$ and $n_s=0.986$).}
    \label{dndm}
    \end{figure}

In Fig.~\ref{dndm}, we plot the halo mass function at several different redshifts found using these two sets of cosmological parameters. At relatively low redshifts ($z\lsim 10$), the differences resulting between these parameter sets is modest. At higher redshifts, however, there are large variations in the predicted number of halos.


The spectrum of photons present at redshift $z$, having been produced previously in dark matter annihilations, is given by
\begin{equation}
\frac{dN_{\gamma}}{dE_{\gamma}}(E_{\gamma},z) = \int_z^{\infty} \frac{dz^{\prime} \,\,R(z^{\prime})}{H(z^{\prime}) (1+z^{\prime})} \frac{dN^{\prime}_{\gamma}}{dE^{\prime}_{\gamma}}(E^{\prime}_{\gamma}) \bigg(\frac{1+z}{1+z^{\prime}}\bigg)^3 \, [1-A_b(z,z^{\prime},E^{\prime}_{\gamma})],
\label{specz}
\end{equation}
where $E^{\prime}_{\gamma} =E_{\gamma}\,(1+z^{\prime})/(1+z)$ is the energy of the gamma ray when it was produced at redshift $z^{\prime}$, and $dN^{\prime}_{\gamma}/dE^{\prime}_{\gamma}(E^{\prime}_{\gamma})$ is the spectrum of photons produced per dark matter annihilation. $A_b(z,z^{\prime},E^{\prime}_{\gamma})$ is the fraction of photons which is absorbed between redshift $z^{\prime}$ and $z$, which we will return to later. The factor of $H(z^{\prime}) (1+z^{\prime})$ in the denominator converts $R(z^{\prime})$ from a number of annihilations per volume, per time, into a number of annihilations per volume, per redshift. This factor can be written as
\begin{equation}
\frac{1}{H(z^{\prime}) (1+z^{\prime})} = \frac{1}{H_0 [\Omega_M \, (1+z^{\prime})^3 +\Omega_{\Lambda}]^{1/2}\, (1+z^{\prime})} \approx \frac{1}{H_0 \Omega^{1/2}_{M}\,(1+z^{\prime})^{5/2}},
\end{equation}
where the last step in the above expression is valid during the matter dominated era, for which $(1+z) >> \Omega_{\Lambda}/\Omega_M$.

The spectrum of gamma rays that is produced through dark matter annihilations, $dN^{\prime}_{\gamma}/dE^{\prime}_{\gamma} (E^{\prime}_{\gamma})$, depends on the characteristics of the dark matter candidate being considered. When dark matter particles annihilate, they can produce a wide variety of particles which fragment and decay into combinations of gamma rays, electrons, neutrinos and protons (and their antimatter counterparts). The gamma rays and electrons (indirectly, through inverse Compton scattering) each contribute to the injected photon spectrum.

The probability per time of a gamma ray scattering with an electron in a hydrogen atom, leading to its ionization, is given by
\begin{equation}
P(E_{\gamma},z) = \sigma_{\gamma e} (E_{\gamma}) \, n_b (1+z)^3 [1-x_{\rm ion}(z)]\, c,
\label{prob}
\end{equation}
where $n_b \approx 2.5 \times 10^{-7}$ cm$^{-3}$ is the current baryon number density, $\sigma_{\gamma e}$ is the Klein-Nishina cross section, and $x_{\rm ion}(z)$ is the fraction of the baryons which is ionized at redshift $z$. Of the energy transfered from the photon in these scatterings, approximately 1/3 goes into the reionzation of atoms~\cite{third}, which induces the following number of ionizations 
\begin{equation}
N_{\rm ion}(E_{\gamma}) \approx \frac{E_{\gamma}}{3} \bigg[\frac{0.76}{0.82}\,\frac{1}{13.5 \, {\rm eV}}+\frac{0.06}{0.82}\,\frac{1}{24.6 \, {\rm eV}} \bigg] \approx 2.4 \times 10^7 \,\bigg(\frac{E_{\gamma}}{1\, {\rm GeV}}\bigg).
\end{equation}
At GeV energies, a large majority of the incident photon's energy is transferred to the scattered electron. At lower energies, the fraction of energy transferred is somewhat reduced. A 1 MeV (10 MeV) photon, for example, loses on average 44\% (68\%) of its energy in such a collision. The remaining lower energy photon, however, will be reasonably likely to scatter again, depositing still more of its energy into ionization and heating. For simplicity, we assume that all of the photon's energy is ultimately transfered in these scatterings.

In the case of a dark matter particle with an electroweak scale mass ($\sim$50-1000 GeV), the majority of the energy in gamma rays is carried away by photons with $\sim$100 MeV or more energy each. A photon with an energy of 1 GeV has a cross section with electrons of only $\sim 10^{-27}$ cm$^2$, however, which leads to a $\sim$0.03\% chance of scattering per billion years (at $z\sim 10$). Highly energetic photons are very inefficient ionizers of gas.

In addition to gamma rays, however, dark matter annihilations can also produce high energy electrons. Such electrons transfer their energy to low energy photons through inverse Compton scattering at a rate of
\begin{equation}
\frac{dE_e}{dt} = \frac{4}{3} \, \sigma_T \, \rho_{\rm rad}\, c \,\bigg(\frac{E_e}{m_e}\bigg)^2,
\end{equation}
where $\rho_{\rm rad}$ is the energy density of radiation and $\sigma_T$ is the Thompson cross section. By scattering off of photons in the cosmic microwave background (CMB), an electron will lose energy at a rate of
\begin{eqnarray}
\frac{dE_e}{dt} &=& \frac{4}{3} \, \sigma_T \, \rho_{\rm CMB}(z=0)\, (1+z)^4 \, c \,\bigg(\frac{E_e}{m_e}\bigg)^2 , \nonumber \\
&\approx & 2.3 \times 10^{-17} \, {\rm GeV}/{\rm s} \,\,\, \,(1+z)^4 \bigg(\frac{E_e}{1 \,{\rm GeV}}\bigg)^2.
\end{eqnarray}
As a result of this process, a 1 GeV electron at redshift $z=6$ will lose 99\% of its energy to the CMB in approximately 50 million years (and 99.9\% of its energy in 500 million years).  At higher redshifts, the transfer of energy is even more rapid. Thus the vast majority of the energy that is deposited via dark matter annihilations into electrons gets almost immediately transfered into lower energy photons.

\begin{figure}[thbp]
    \sfig{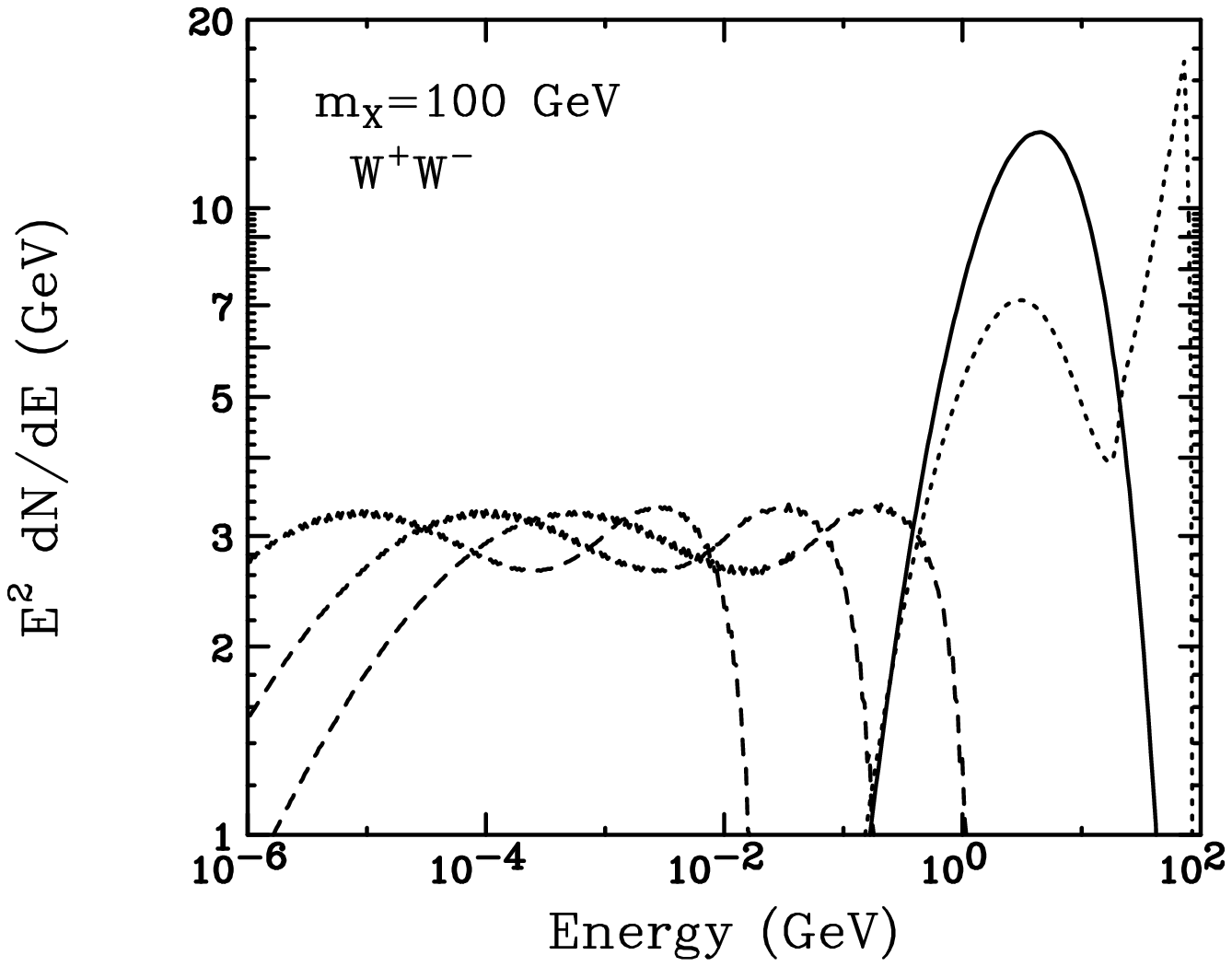}{0.49\columnwidth}
    \sfig{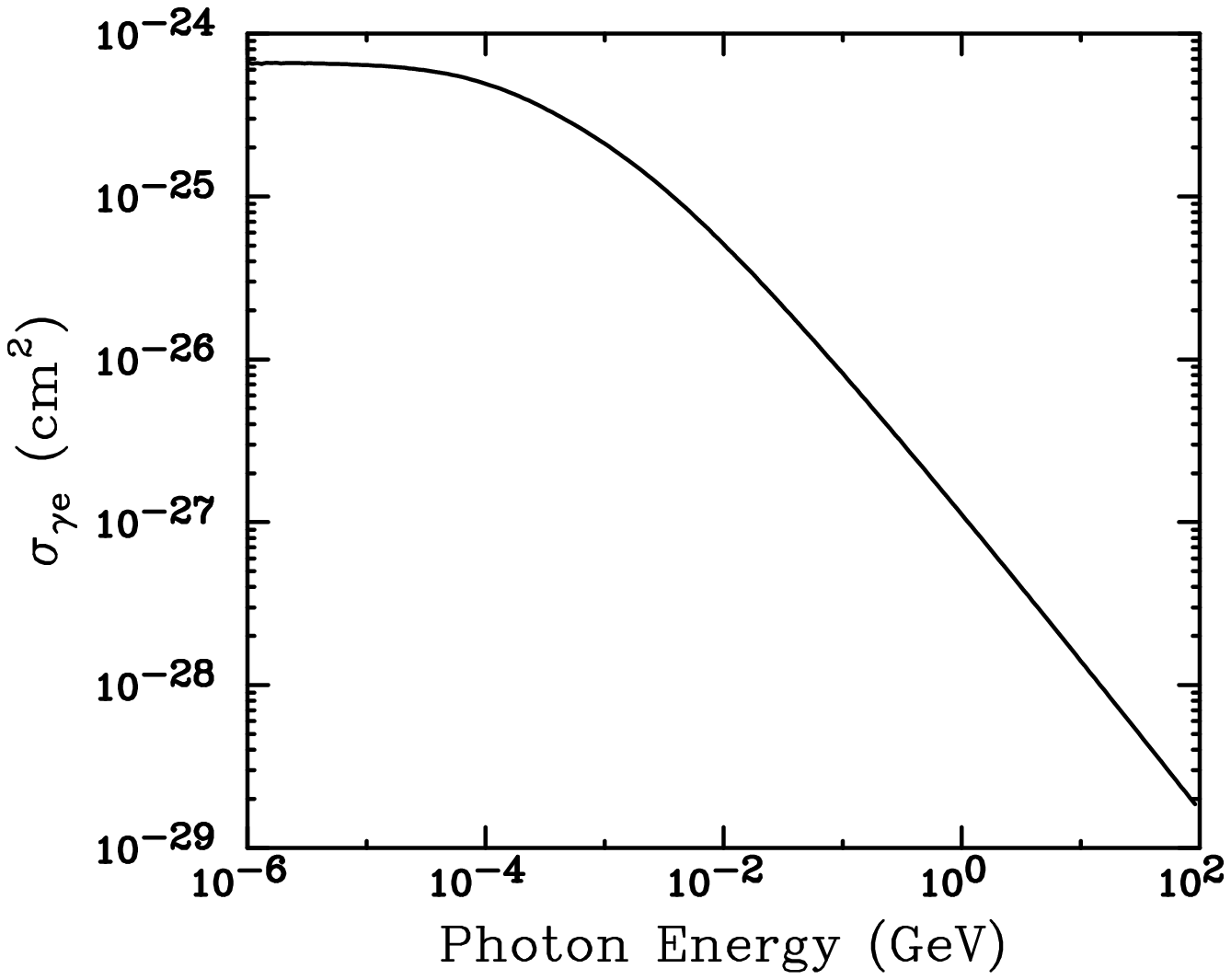}{0.49\columnwidth}
    \caption{In the left frame, we plot the spectrum of prompt gamma rays (solid) and electrons (dots) from the annihilation of 100 GeV dark matter particles to $W^+ W^-$. We also plot, as dashed lines, the spectrum of inverse Compton photons which results from those electrons scattering with the cosmic microwave background (for redshifts of $z=$0, 10 and 60, from left to right). In the right frame, the Klein-Nishina cross section is shown as a function of photon energy. Due to the rapidly falling cross section, the inverse Compton photons are far more efficient at reionizing the universe than higher energy prompt gamma rays.}
    \label{IC}
    \end{figure}

An inverse Compton scattering between an energetic electron and a CMB photon results in a photon with an average energy of 
\begin{eqnarray}
E_{\rm IC} &=& \frac{4}{3} \, \bigg(\frac{E_e}{m_e}\bigg)^2 \, E_{\rm CMB} \nonumber \\
           &\approx& 3.2 \times 10^{-4} \, {\rm GeV} \,\, \, (1+z) \, \bigg(\frac{E_e}{10 \, {\rm GeV}}\bigg)^2.
\end{eqnarray}
As the Klein-Nishina cross section is more than two orders of magnitude larger at the typical energies of these inverse Compton photons ($\sim$$10^{-3}$\,GeV) than at the energies of prompt photons ($\sim$\,GeV), we conclude that electrons from dark matter annihilations provide us with a far more efficient mechanism for reionizing the universe. This is illustrated in Fig.~\ref{IC}, where we plot the spectrum of prompt gamma rays (solid) and electrons (dots) from the annihilations of a 100 GeV dark matter particle which annihilates to $W^+ W^-$ (we have used PYTHIA~\cite{pythia} to calculate these spectra).  At lower energies, we plot as dashed lines the spectrum of inverse Compton photons which results from those electrons scattering with the CMB (for three different redshifts). For further details regarding the spectrum resulting from inverse Compton scattering, see Ref.~\cite{lightman}. In the right frame of the figure, the Klein-Nishina cross section is shown as a function of photon energy. The inverse Compton photons naturally fall within or near the range in which the cross section is approximately equal to the Thompson cross section. Prompt photons, in contrast, typically have much greater energy and thus a much smaller cross section.

Combining these expressions, we arrive at the number of ionizations induced per volume, per time, at a redshift $z$,
\begin{eqnarray}
I(z) &=& \int^{m_X}_0 dE_{\gamma} \, \frac{dN_{\gamma}}{dE_{\gamma}}(E_{\gamma},z) \, P(E_{\gamma},z) \, N_{\rm ion}(E_{\gamma})\nonumber \\
&=& \frac{[1-x_{\rm{ion}}(z)]\,(1+z)^6\, n_b \, c \, (2.4 \times 10^7 \, {\rm GeV}^{-1})}{H_0 \Omega_M^{1/2}}\,\frac{\sigv}{2 m^2_X} \int^{m_X}_0 dE_{\gamma}   \sigma_{\gamma e} (E_{\gamma})   E_{\gamma} \,    \nonumber \\
  &\times& \int_z^{\infty} dz^{\prime} \,\frac{A_b(z,z^{\prime},E^{\prime}_{\gamma})}{(1+z^{\prime})^{5/2}}    \frac{dN^{\prime}_{\gamma}}{dE^{\prime}_{\gamma}}(E^{\prime}_{\gamma}) \, \int^{\infty}_{M_{\rm min}} \frac{dn}{dM}(M,z^{\prime}) dM  \int \rho^2(r,M) 4 \pi r^2 dr.
\end{eqnarray}

Where $A_b(z,z^{\prime},E^{\prime}_{\gamma})$ denotes the fraction of photons which are absorbed between redshifts $z^{\prime}$ and $z$,
\begin{equation}
A_b(z,z^{\prime},E^{\prime}_{\gamma}) \approx \exp\bigg[\int_{z}^{z^{\prime}} \frac{dz^{\prime \prime} \sigma_{\gamma e}(E^{\prime \prime}_{\gamma}) \, n_b \, (1+z^{\prime \prime})^3 \, c}{H_0 \, \Omega_M^{1/2} \, (1+z^{\prime \prime})^{5/2}}\bigg].
\end{equation}
In addition, we consider the competing effect of recombination which, neglecting significant baryon clumping, occurs at a rate per volume, per time, given by
\begin{equation}
R_c(z) \approx n_b^2 \, x^2_{\rm ion}(z)\, (1+z)^6 \bigg[\frac{0.76}{0.82}\alpha_{\rm H}(z) +\frac{0.06}{0.82}\alpha_{\rm He}(z)  \bigg], 
\end{equation}
where~\cite{alpha}
\begin{eqnarray}
\alpha_{\rm H}(z) &\approx& 3.75 \times 10^{-13} \, {\rm cm}^3/{\rm s} \,\, \bigg(\frac{T(z)}{1 \,{\rm eV}}\bigg)^{-0.724} \nonumber \\
\alpha_{\rm He}(z) &\approx& 3.93 \times 10^{-13} \, {\rm cm}^3/{\rm s} \,\, \bigg(\frac{T(z)}{1 \,{\rm eV}}\bigg)^{-0.635}.
\end{eqnarray}
In the absence of heating, we estimate the temperature of the gas to vary with redshift as $T(z) \approx 5.3 \times 10^{-3} \, [(1+z)/61]^2$ eV. In addition, however, the energy released in dark matter annihilations can heat the gas and suppress the rate of recombination. In our calculation, we assume that one third of the energy transfered from dark matter annihilation products into atoms via electron scattering goes into gas heating~\cite{third}. 

The contributions from ionization and recombination collectively lead to the following rate of change in the fraction of ionized baryons
\begin{equation}
\frac{dx_{\rm ion}}{dt}(z) = \frac{I(z)-R_c(z)}{n_b(1+z)^3},
\end{equation}
or equivalently,
\begin{equation}
\frac{dx_{\rm ion}}{dz}(z) \approx \frac{I(z)-R_c(z)}{-n_b(1+z)^{11/2} \,H_0 \, \Omega^{1/2}_M}.
\end{equation}

\begin{figure}[thbp]
    \sfig{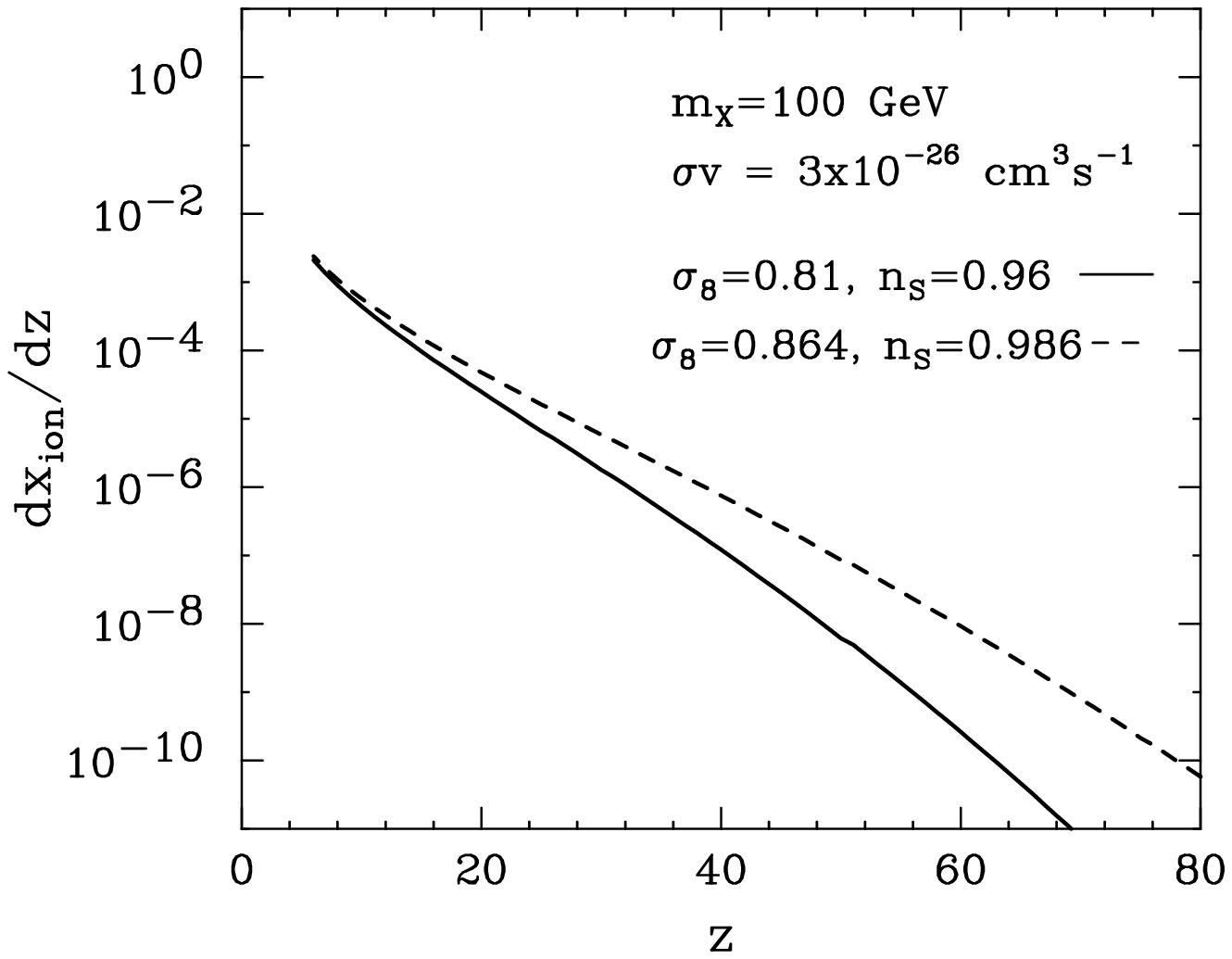}{0.49\columnwidth}
    \sfig{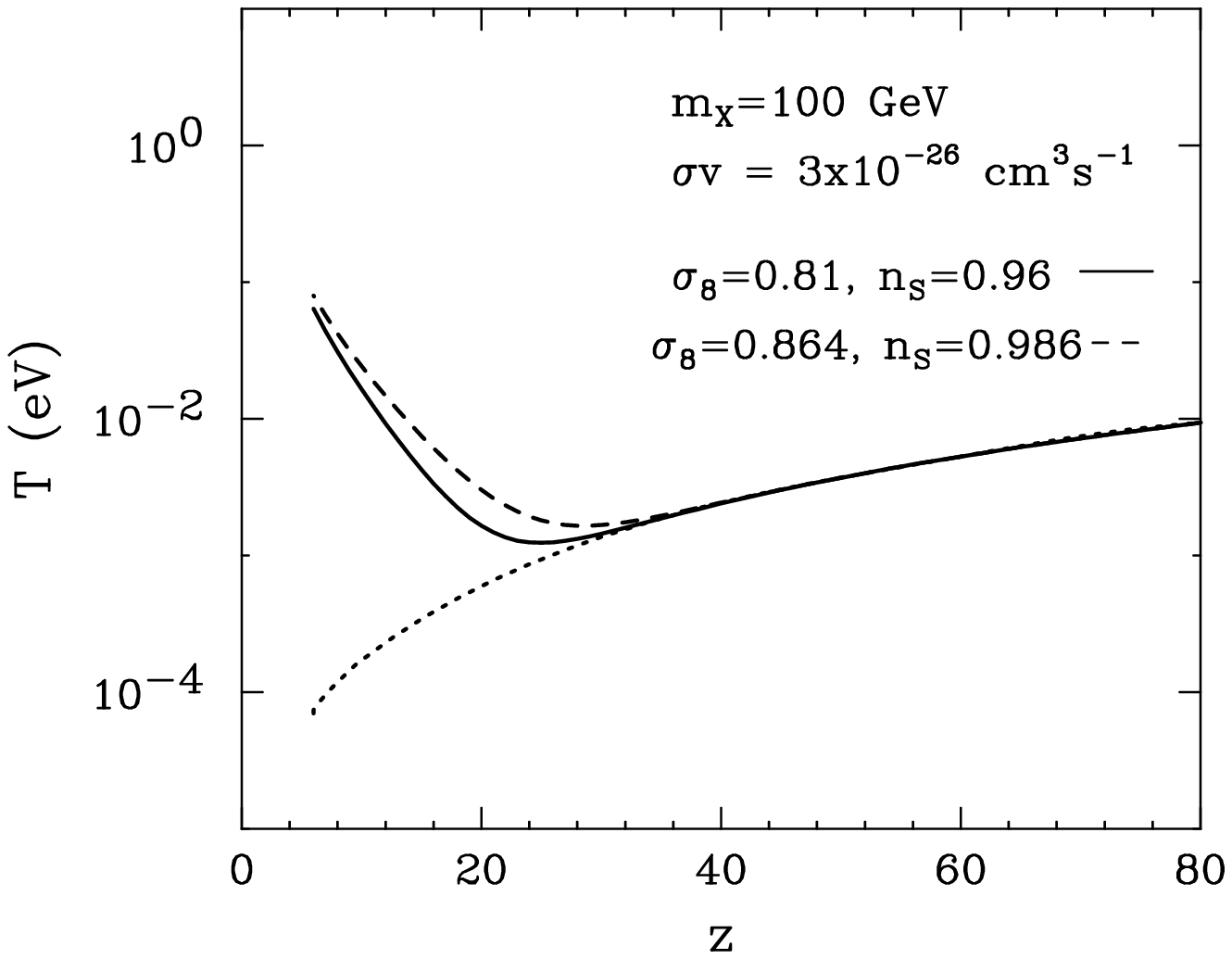}{0.49\columnwidth}\\
\sfig{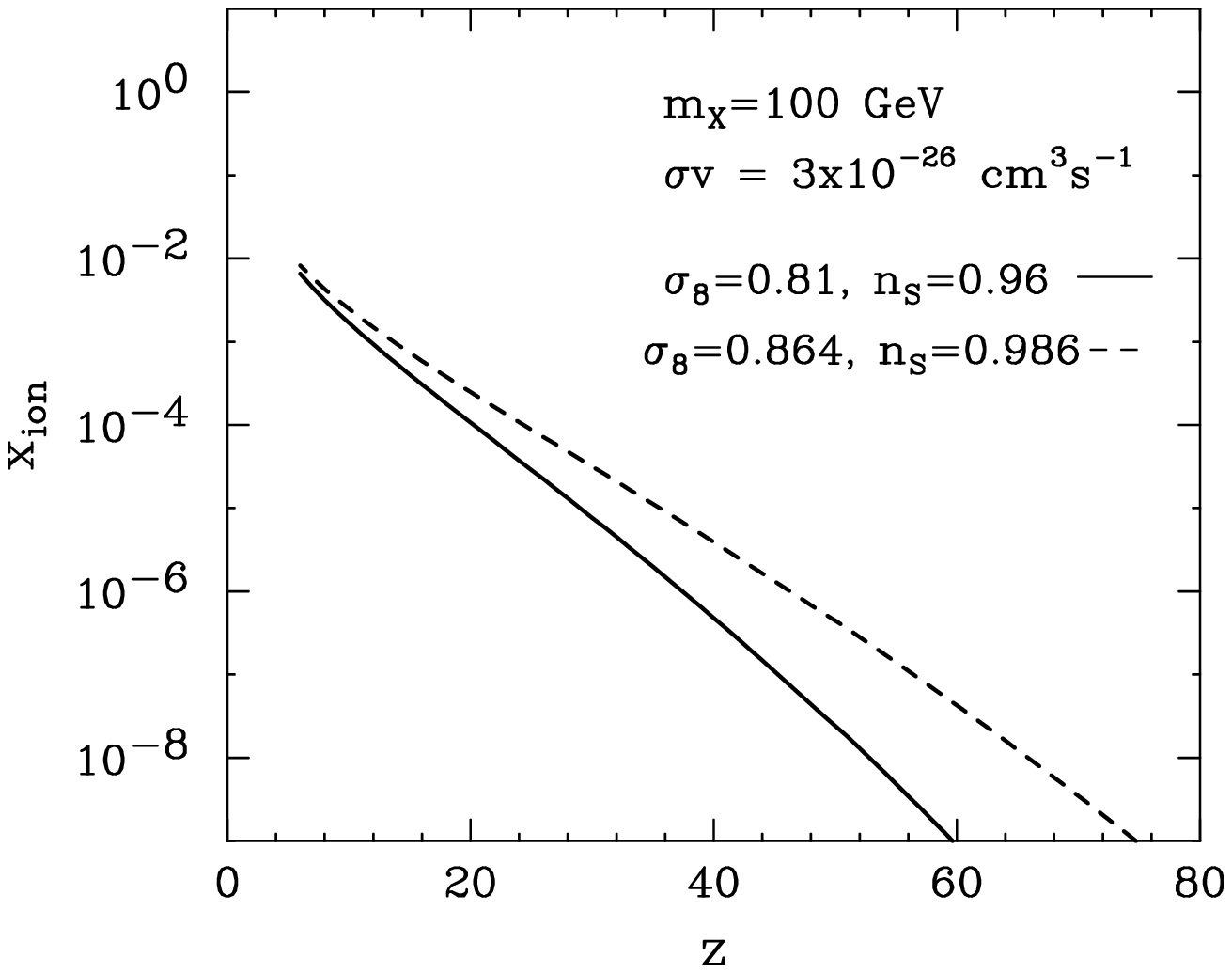}{0.49\columnwidth}
    \caption{The rate of change of the fraction of ionized baryons (upper left), the temperature of gas (upper right), and the fraction of ionized baryons (lower), as a function of redshift. Here, we have considered a 100 GeV dark matter particle which annihilates to $W^+W^-$ with a cross section of $\sigv = 3 \times 10^{-26}$ cm$^3$/s. We show results using two sets of cosmological parameters ($n_s = 0.96, \sigma_8 = 0.812$ and $n_s = 0.986, \sigma_8 = 0.864$). In the upper right frame, the dotted line denotes the evolution of the gas temperture without heating from dark matter annihilations.}
    \label{xionrate}
    \end{figure}

In Fig.~\ref{xionrate}, we show the effect on the ionization history of the universe of the annihilations of a 100 GeV dark matter particle which annihilates to $W^+W^-$ with a cross section of $\sigv = 3 \times 10^{-26}$ cm$^3$/s (the value for a typical thermal relic). In the upper left frame, we show the rate at which the fraction of ionized baryons changes as a function of redshift as a result of dark matter annihilations. In the upper right frame, we plot the evolution of the gas temperature. In the lower frame, the ionized fraction is shown. In each frame, we show results for two sets of cosmological parameters ($n_s = 0.96, \sigma_8 = 0.812$ and $n_s = 0.986, \sigma_8 = 0.864$). In each case, we have also adopted a minimum halo mass of $M_{\rm min} = 10^{-8}\, M_{\odot}$.

From the ionized fraction of baryons over cosmic history, we can calculate the resulting Thompson optical depth of the universe,
\begin{equation}
\tau =  n_b \, \sigma_T \bigg[ -\frac{0.88}{0.82}\int^3_0 dz \frac{dt}{dz}(1+z)^3 - \int^6_3 dz \frac{dt}{dz}(1+z)^3 - \int^{\infty}_6 dz \frac{dt}{dz} (1+z)^3 x_{\rm ion}(z) \bigg].
\label{taueq}
\end{equation}
In this expression, the first (second) term accounts for the contribution since $z=3$ (between $3<z<6$), and assumes the helium to be doubly (singly) ionized~\cite{doubly,doubly2}. The last term describes the contribution prior to $z=6$.

Our empirical knowledge regarding the ionization history of the universe consists of essentially two observations.  Firstly, the lack of significant Ly$\alpha$ absorption observed in the spectra of quasars lead us to conclude that the universe has been highly ionized since a redshift of $z\approx 6$~\cite{gunn}. This is why we have assumed complete ionization in the first two terms in Eq.~\ref{taueq}. Secondly, WMAP has measured the Thompson optical depth of the universe to be $\tau \approx 0.087 \pm 0.017$~\cite{wmap}. The contribution to this quantity from a fully ionized universe since $z=6$ is approximately 0.04, considerably less than the total measured quantity. This forces us to conclude that sources prior to $z=6$ have contributed approximately half of the total optical depth of the universe.

We can use the values of $x_{\rm ion}$ shown in the lower frame of Fig.~\ref{xionrate} to calculate the total optical depth of the universe and compare it to the value measured by WMAP. For this choice of dark matter mass, cross section, and dominant annihilation channel, dark matter annihilations lead to on the order of 1\% of all baryons being ionized by $z=6$, and to a total contribution to the optical depth prior to $z=6$ of $\delta \tau \approx 0.00029$ or 0.00043, for the two sets of cosmological parameters considered. These values are far too small to account for the optical depth observed by WMAP.

\begin{figure}[thbp]
    \sfig{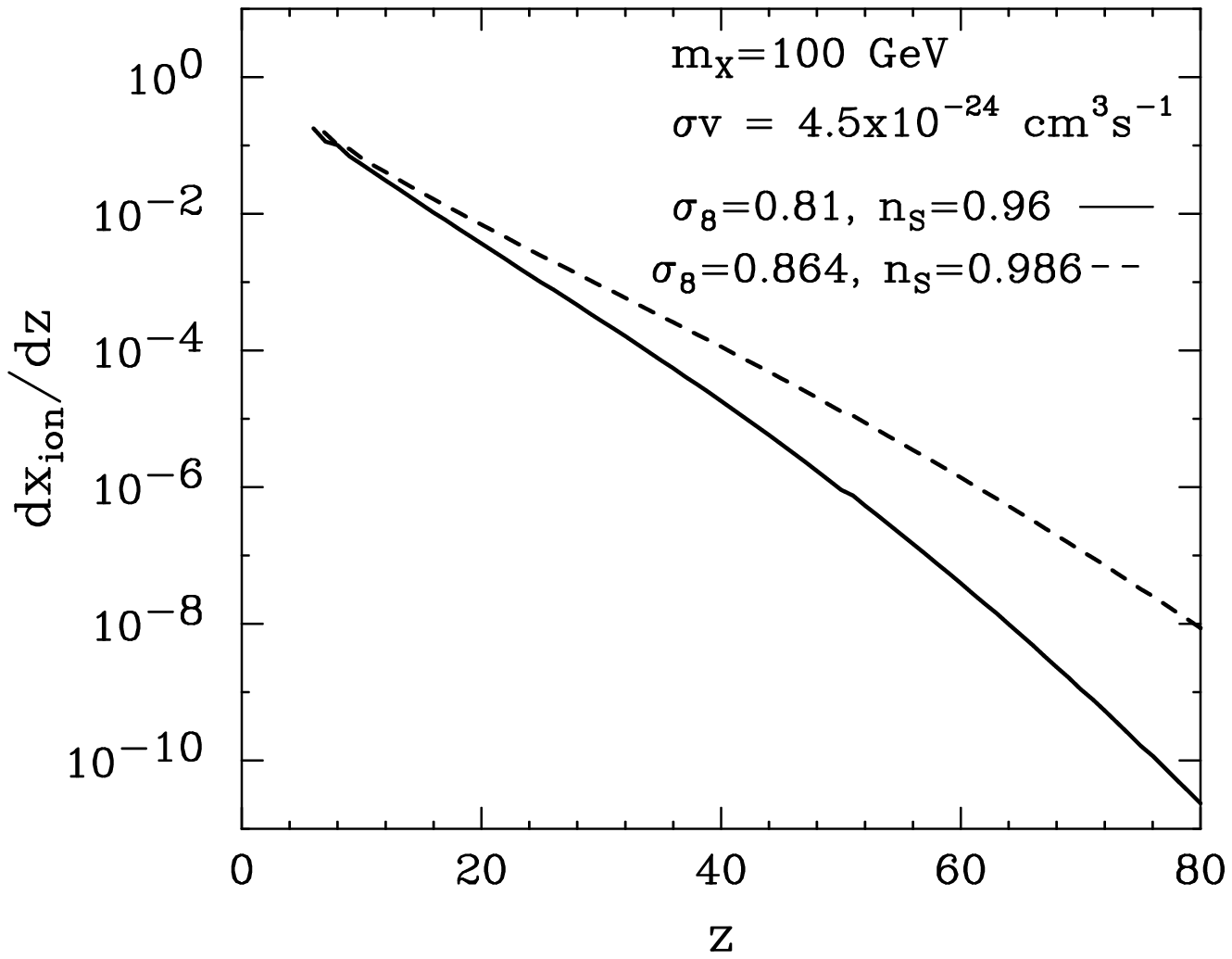}{0.49\columnwidth}
    \sfig{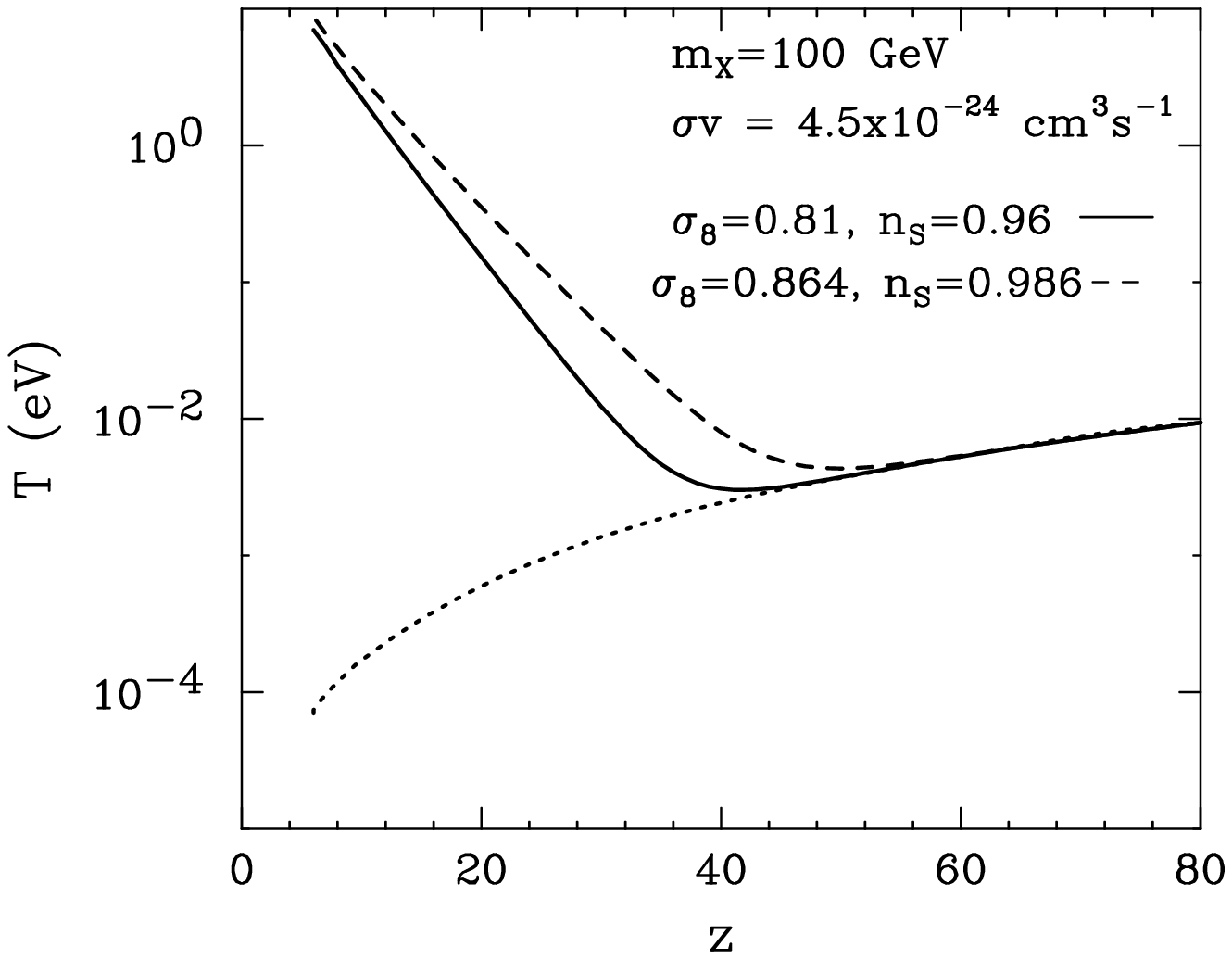}{0.49\columnwidth}\\
\sfig{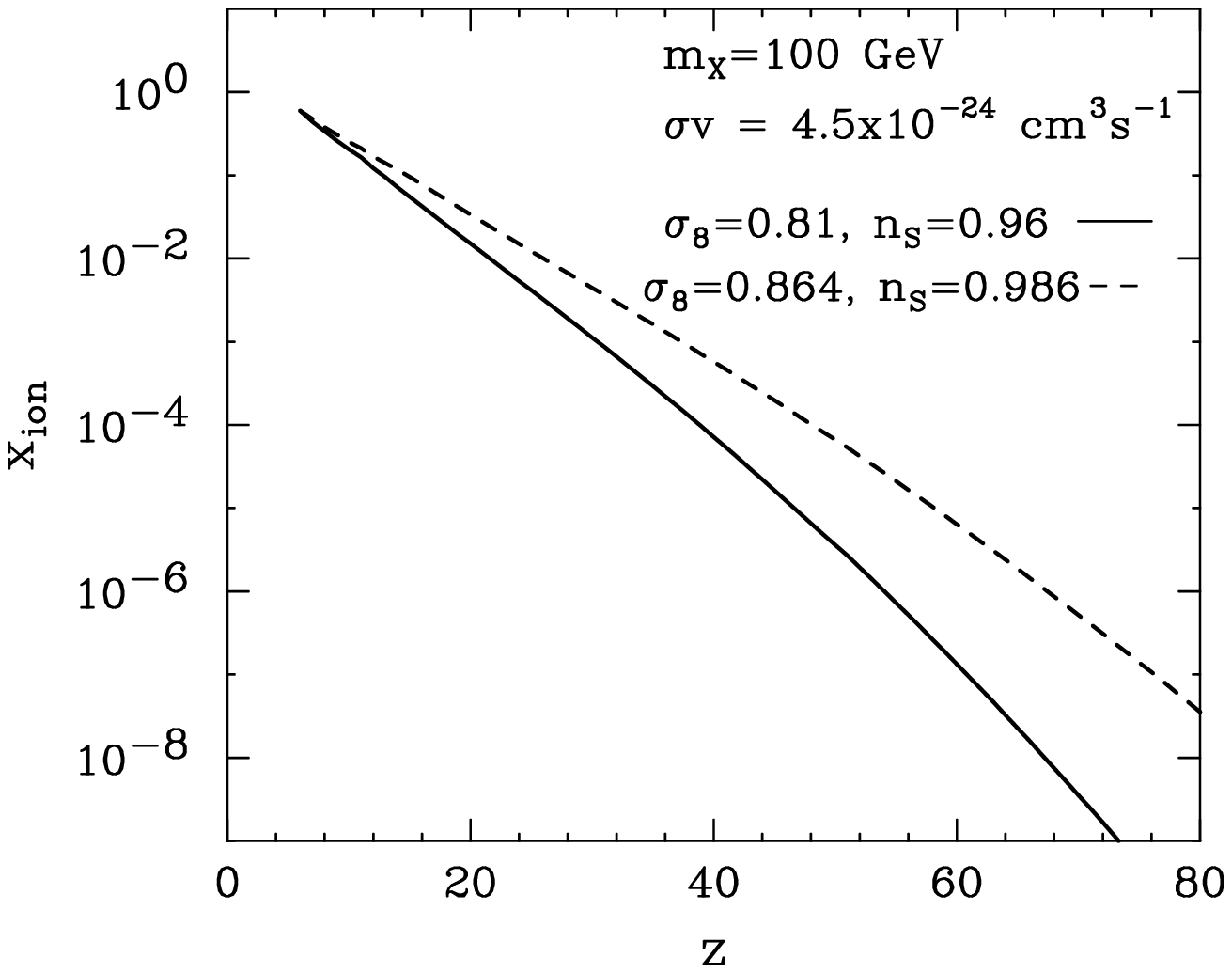}{0.49\columnwidth}
    \caption{The same as shown in Fig.~\ref{xionrate}, but for the case of a 100 GeV dark matter particle which annihilates to $W^+W^-$ with a cross section of $4.5\times 10^{-24}$ cm$^3$/s (a wino-like neutralino, for example). Dark matter annihilations in this model lead to nearly total ionization by $z\approx 6$, and constitute the primary source the optical depth as measured by WMAP.}
    \label{xionratewino}
    \end{figure}

For dark matter annihilations to account for the total opacity observed by WMAP, they must reionize the universe at a considerably higher rate than found in this first example. This could be potentially accomplished in a number of ways, the most straightforward being to simply increase the annihilation cross section and corresponding annihilation rate of the dark matter particle. Consider, for example, dark matter in the form of a wino-like neutralino. Such dark matter candidates, which appear naturally in models of anomaly mediated supersymmetry breaking, have larger annihilation cross sections than that expected from a typical thermal relic. A 100 GeV wino, for example, has an annihilation cross section to $W^+ W^-$ of approximately $\sigv \approx 4.5\times 10^{-24}$ cm$^3$/s. If winos are produced in the early universe through a non-thermal mechanism~\cite{wino,winononthermal}, they can constitute the measured cosmological dark matter abundance despite their high annihilation rate.

\begin{figure}[thbp]
    \sfig{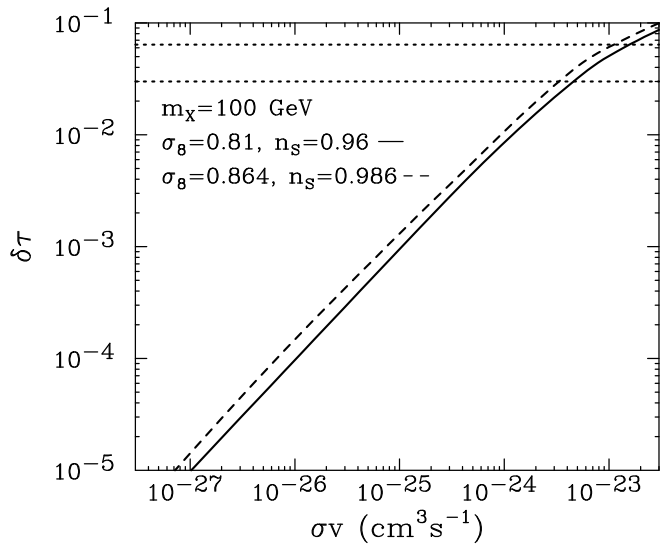}{0.49\columnwidth}
    \caption{The contribution to the optical depth of the universe (over $z > 6$) from dark matter annihilations. Here we have considered a 100 GeV dark matter particle which annihilates to $W^+W^-$. The horizontal dotted lines denotes the range of values measured by WMAP, $\delta \tau \approx 0.047 \pm 0.017$. A relatively light (100-200 GeV) wino-like neutralino would naturally lead to an optical depth consistent with this measurement.}
    \label{tau}
    \end{figure}

In Fig.~\ref{xionratewino}, we show the ionization history of the universe resulting from a 100 GeV, wino-like dark matter particle. For this dark matter candidate, we find that nearly full reionization is reached by $z\approx 6$. In Fig.~\ref{tau}, we show the contribution to the optical depth prior to $z=6$ resulting from a 100 GeV dark matter particle annihilating to $W^+W^-$, as a function of its annihilation cross section. From this figure, we see that such a dark matter particle can generate the total measured optical depth if it possesses a cross section of approximately $\sim (3-10)\times 10^{-24}$ cm$^3$/s. Dark matter in the form of a $\sim100$ GeV wino is thus predicted to lead to an optical depth similar to that measured by WMAP, even without any significant contributions from quasars or early stars.

\begin{figure}[thbp]
    \sfig{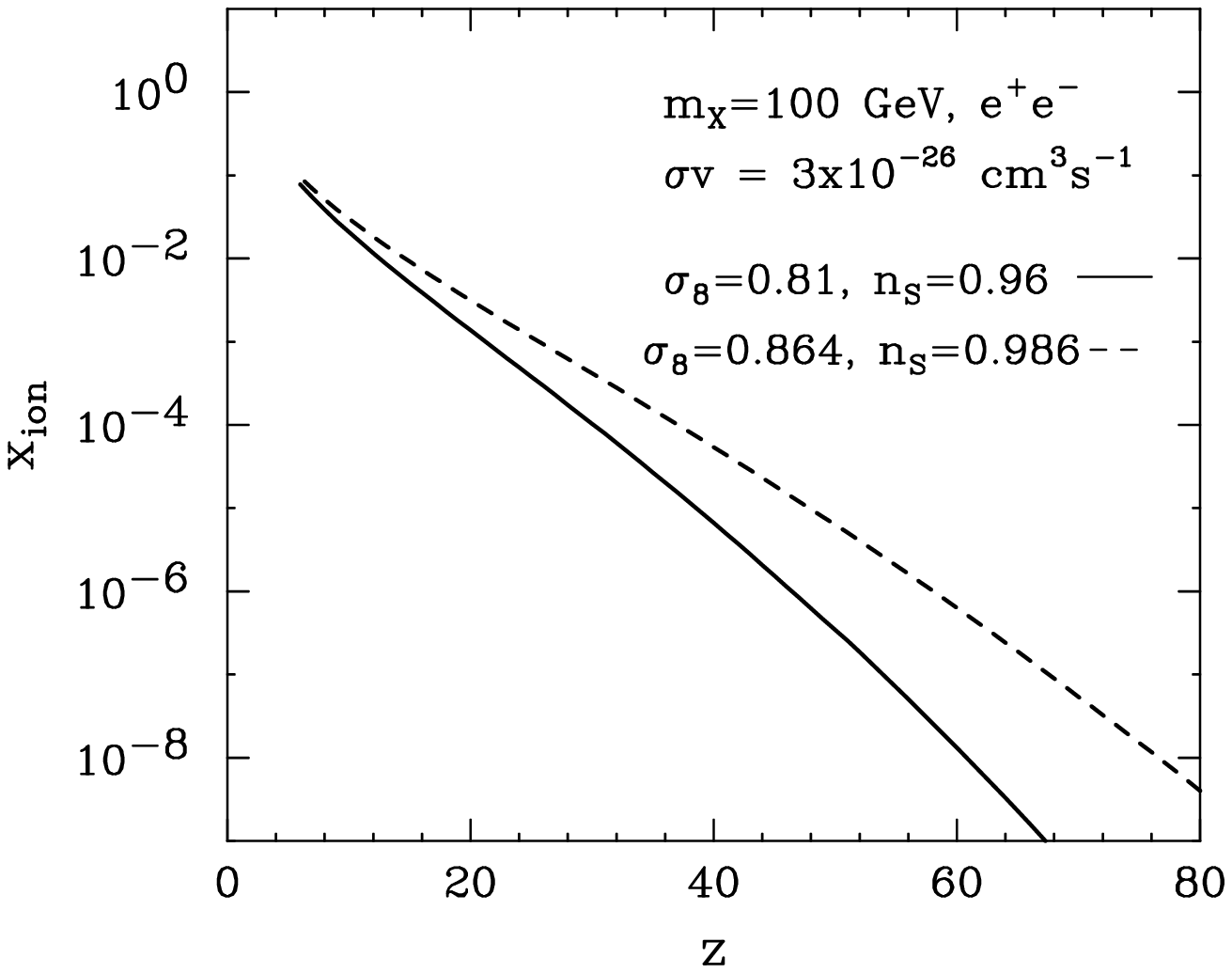}{0.49\columnwidth}
    \sfig{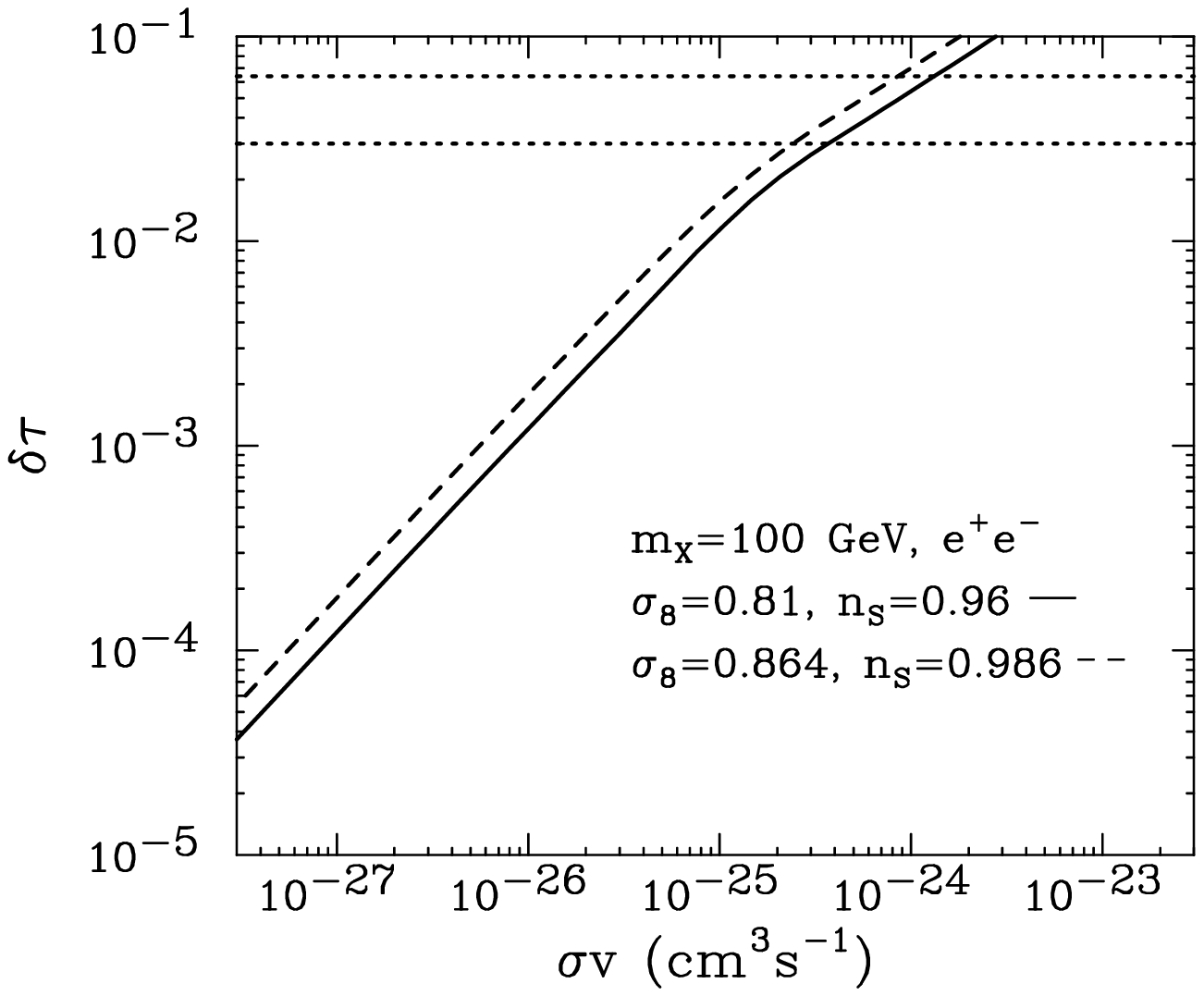}{0.49\columnwidth}\\
    \sfig{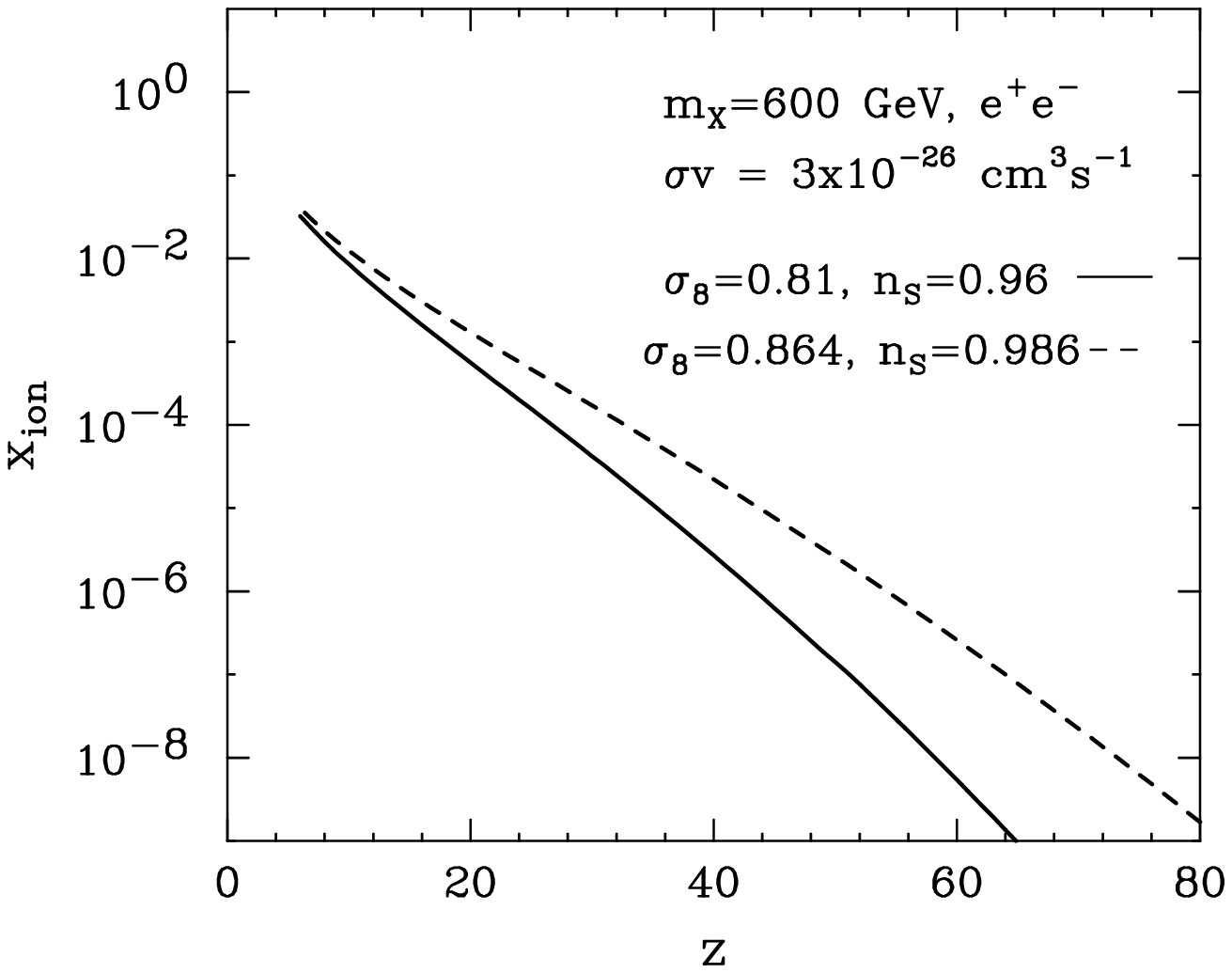}{0.49\columnwidth}
    \sfig{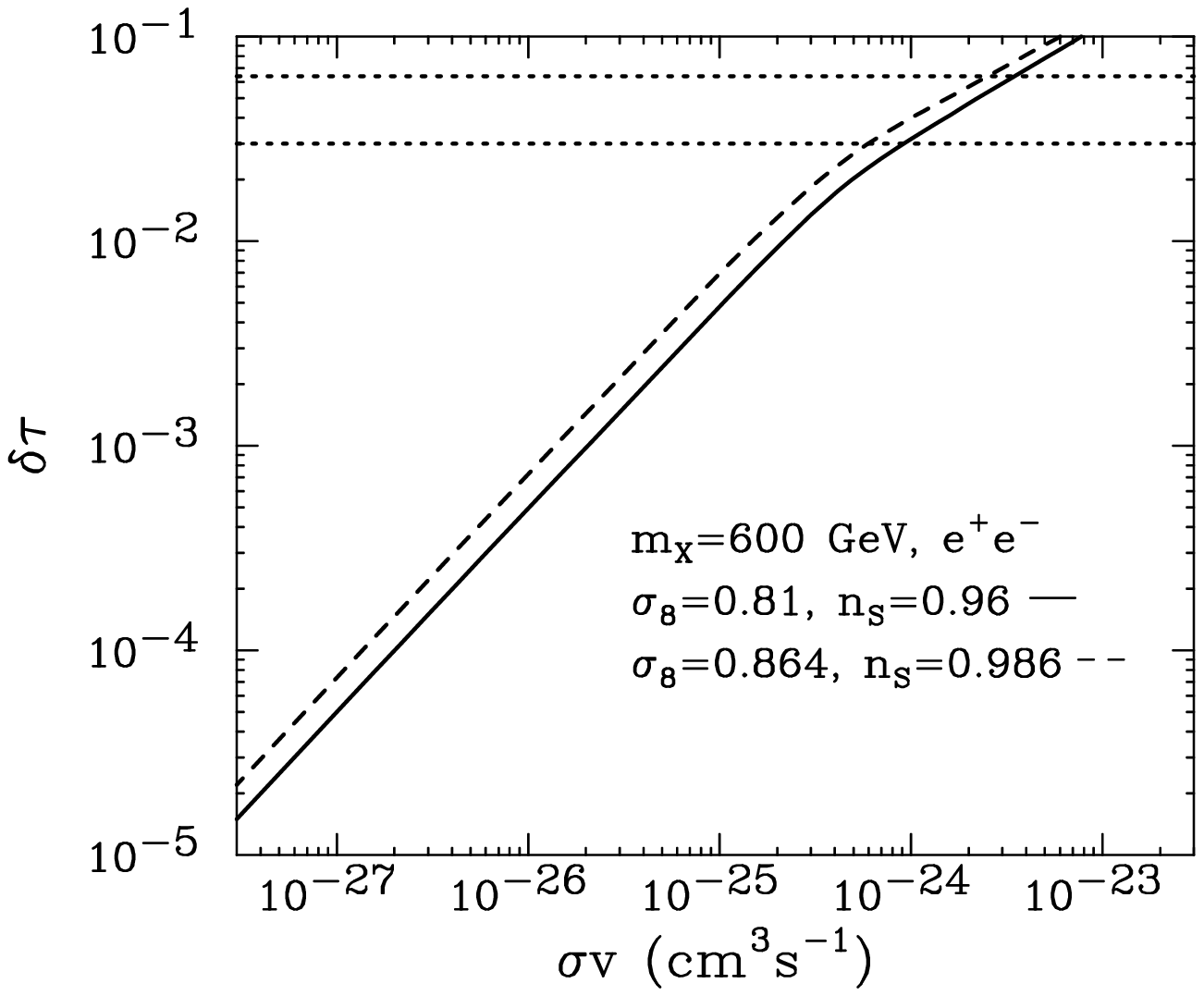}{0.49\columnwidth}
    \caption{In the left frames, we show the fraction of ionized baryons, as a function of redshift. We have considered 100 GeV (upper) and 600 GeV (lower) dark matter particles which annihilate to $e^+e^-$. In the left frames, we have used a cross section of $\sigv = 3 \times 10^{-26}$ cm$^3$/s. In the right frames, we show the contribution to the optical depth of the universe (over $z > 6$) from dark matter annihilations to $e^+ e^-$, as a function of the annihilation cross section. The horizontal dotted lines denotes the range of values measured by WMAP, $\delta \tau \approx 0.047 \pm 0.017$. In each frame, we show results using two sets of cosmological parameters ($n_s = 0.96, \sigma_8 = 0.812$ and $n_s = 0.986, \sigma_8 = 0.864$).}
    \label{xionrateee}
    \end{figure}

Another way to potentially increase the contribution to the optical depth of the universe is to consider dark matter particles which annihilate largely to electron-positron pairs (or $\mu^+ \mu^-$ or $\tau^+ \tau^-$), which deposit a larger fraction of their energy into inverse Compton photons. In Fig.~\ref{xionrateee}, we show the ionization history and optical depth resulting from dark matter which annihilates to $e^+ e^-$. Comparing the results for a 100 GeV dark matter particle to those found in Fig.~\ref{tau}, we find that the $e^+ e^-$ annihilation channel is approximately an order of magnitude more efficient in reionizing gas than annihilations to $W^+ W^-$. In the lower frames of Fig.~\ref{xionrateee}, we also show the results for a 600 GeV dark matter particle which annihilations to $e^+ e^-$.

\section{Implications of PAMELA and ATIC For Reionization}

Recently, the PAMELA~\cite{PAMELA} and ATIC~\cite{ATIC} collaborations have announced observations of surprisingly large fluxes of high energy electrons and positrons in the cosmic ray spectrum. These observations appear to imply the presence of a relatively local source of energetic electron-positron pairs. Although the origin of these particles remains unknown, a nearby pulsar~\cite{pulsars,pulsars2} and dark matter annihilations~\cite{pamdm1,pamdm2} have each been proposed as possible sources. 

If annihilating dark matter particles are to explain the signals of PAMELA and/or ATIC, they must have some rather specific properties, however. In particular, they must annihilate at a rate considerably larger than expected for a simple thermal relic distributed smoothly throughout the Galactic Halo. Furthermore, to reproduce the very hard spectra observed by these experiments, the dark matter particles must annihilate largely to $e^+ e^-$ or other charged leptons (for examples of models designed to possess this feature, see Refs.~\cite{leptons1,leptons2,leptons3,leptons4,leptons5,leptons6}). 

Dark matter particles with large annihilations and/or which annihilate to charged leptons are, or course, precisely what we have shown to be needed if dark matter is to play a significant role in the reionization of the universe. More specifically, to produce the PAMELA excess (neglecting ATIC for the moment) with 100 GeV dark matter particles which annihilate to $e^+ e^-$, an annihilation cross section of approximately $\sigv \sim (7.2-69) \times 10^{-26}$ cm$^3$/s is required (assuming no large boost factors from local inhomogeneities in the dark matter distribution)~\cite{pamdm1}. From the upper right frame of Fig.~\ref{xionrateee}, we see that this range of cross sections can naturally lead to the observed optical depth of the universe. Fitting the ATIC signal to 600 GeV dark matter particles, we are forced to require an annihilation cross section to $e^+ e^-$ of approximately $\sigv \sim (1-2) \times 10^{-24}$ cm$^3$/s (see, for example, Ref.~\cite{kkdm}). Again, from the lower frame of Fig.~\ref{xionrateee} we see that this range of cross sections naturally produces the desired amount of reionization. This conclusion holds regardless of whether the large annihilation cross section is accommodated by a non-thermal production mechanism, or results from Sommerfeld-type enhancements (assuming saturation occurs at velocities not far below typical velocities of the Milky Way)~\cite{sommerfeld1,sommerfeld2,sommerfeld3}. We thus reach the very interesting conclusion that dark matter particles with the characteristics required to explain PAMELA and/or ATIC observations also invariably lead to very significant contributions to reionization.

\section{Summary and Conclusions}

In this article, we have calculated the contribution to the reionization of gas that results from annihilating dark matter particles. The primary mechanism for ionization is the production of gamma rays through the inverse Compton scattering of high energy electrons, which are themselves products of dark matter annihilations. The inverse Compton photons have a much larger cross section for scattering with electrons, and are thus more efficient in ionizing and heating gas, than higher energy prompt photons. 

Although dark matter particles with typical thermal annihilation cross sections ($\sigv \sim 3 \times 10^{-26}$ cm$^3$/s) only produce about 1-10\% of the ionization rate required to explain the optical depth of the universe observed by WMAP, dark matter particles with larger cross sections can completely reionize the universe by $z\sim 6$ and provide the total observed optical depth. Dark matter particles which annihilate directly to $e^+ e^-$ or other charged leptons lead to a higher ionization rate than particles which annihilate to gauge bosons or quarks. 

Intriguingly, we note that dark matter candidates capable of producing the cosmic ray signals observed by PAMELA and/or ATIC generally possess cross sections and dominant annihilation channels which will lead to the significant ionization of gas in the universe by $z\sim 6$. If either or both of these observations are, in fact, products of dark matter annihilations, then dark matter should be expected to play a major role in the reionization history of the universe.

\bigskip

We would like to thank Nick Gnedin and Aravind Natarajan for very helpful discussions. This work has been supported by the US Department of Energy, including grant DE-FG02-95ER40896, and by NASA grant NAG5-10842.

\end{document}